\documentclass[12pt]{iopart}
\usepackage{amsfonts}
\usepackage{amssymb}
\usepackage{amscd}
\def\D{\hbox{D\kern-.73em\raise.25ex\hbox{-}\raise-.25ex\hbox{ }}}
 \def\d{\hbox{d\kern-.33em\raise.75ex\hbox{-}\raise-.75ex\hbox{}}}

\def\GGG{\frak G }
\def\gr3{\GGG\,(\SSS_3)}

\def\gr2{\GGG\,(\SSS_2)}

\def\SSS{\frak S}

\def\ed{\end{document}}
\def\beq{\begin{equation}}
\def\eeq{\end{equation}}
\def\bea{\begin{eqnarray}}
\def\eea{\end{eqnarray}}
\def\ba{\begin{array}}
\def\ea{\end{array}}
\def\bi{\begin{itemize}}
\def\ei{\end{itemize}}

\def\nn{\nonumber}

\newcommand{\bp}{\noindent\begin{minipage}[c]}
\newcommand{\ep}{\end{minipage}}

\begin{document}

\title{Power series everywhere convergent on ${\mathbb R}$ and all ${\mathbb
Q}_p $}

\author{Branko G. Dragovich }
\address{ Institute of Physics,  P.O.
Box 57, 11001 Belgrade, Yugoslavia}

\date{}


\begin{abstract}{Power series are introduced that are simultaneously
convergent for all real and $p$-adic numbers. Our expansions are
in some aspects similar to those of exponential, trigonometric,
and hyperbolic functions. Starting from these series and using
their factorial structure new and summable series with rational
sums are obtained. For arguments $x\in {\mathbb Q}$ adeles of
series are constructed. Possible applications at the Planck scale
are also considered.}
\end{abstract}

\section{ INTRODUCTION }

The field of rational numbers is of central importance in physics
and mathematics. It is well known that all results of measurements
belong to ${\mathbb Q}$, i.e. that the irrational numbers cannot
be measured. From a mathematical point of view ${\mathbb Q}$ is
the simplest infinite number field. Completion of ${\mathbb Q}$
with respect to the absolute value gives the field of real numbers
${\mathbb R}$. Algebraic closure of ${\mathbb R}$ leads to the
field of complex numbers ${\mathbb C}$. Although experimental
results are given in ${\mathbb Q}$, theoretical models are usually
constructed over ${\mathbb R}$ or  ${\mathbb C}$. Comparison
between theory and experimental results performs within ${\mathbb
Q}$.

However, it is interesting that in addition to the standard
absolute value there exist $p$-adic norms (valuations) on
${\mathbb Q}$. Completions of ${\mathbb Q}$ with respect to
$p$-adic norms give us the fields of $p$-adic numbers ${\mathbb
Q}_p$  ($p$ = a prime number). There is also a $p$-adic analog of
the complex numbers. According to this similarity between $p$-adic
and real numbers, it is natural to expect that $p$-adic numbers
should also play a significant role in theoretical and
mathematical physics.

Since 1987, $p$-adic numbers have been successfully considered in
string theory \cite{volovich1}, quantum mechanics
\cite{volovich2}, quantum field theory \cite{roth}, and in some
other branches of theoretical \cite{dragovich1,dragovich2} and
mathematical \cite{zelenov} physics. Such new theoretical
constructions are $p$-adic analogs of some models on real (or
complex) numbers.

There has been also a research on various $p$-adic aspects of the
perturbation series \cite{dragovich3}. It is shown that the usual
perturbation series, which are divergent in the real case, are
$p$-adic convergent. Summability of a given series in all but a
finite number of ${\mathbb Q}_p$ may be used for summation of a
divergent counterpart at the rational points.

In order to make a direct connection of $p$-adic models with the
real one it seems to be necessary to have convergence in ${\mathbb
R}$ and all   ${\mathbb Q}_p$ within the common domain of rational
numbers. However, the standard power series of theoretical physics
do not satisfy this property. For example, expansions of functions
$\exp x, \sin x, \cos x, \sinh x,$ and $\cosh x$ are convergent in
the $p$-adic case for $|x|_p < 1$ if $p\neq 2$ and $|x|_2 <
\frac{1}{2}$. As a consequence, there is no $0 \neq x\in {\mathbb
Q}_p$ for which these functions are defined for any $p$.

This paper is devoted  to the power series  that converge
everywhere on ${\mathbb R}$  and everywhere  on ${\mathbb Q}_p$
for every $p$. These analytic functions are simple and suitable
modifications of expansions for exponential, trigonometric, and
hyperbolic functions.

An appropriate mathematical background on $p$-adic numbers and
$p$-adic analysis can be found in Refs.
\cite{schikhof}-\cite{gelfand}.

\section{ EVERYWHERE CONVERGENT SERIES}

In theoretical physics we often encounter a power series \bea
\sum_{n=0}^\infty A_n x^n  \, ,\label{2.1} \eea where $A_n \in
{\mathbb Q}$ and $x\in {\mathbb Q}$. If we take $x\in {\mathbb
Q}_p$ series (\ref{2.1}) may be treated as the $p$-adic one. It is
obvious that the domain of convergence for any of the number
fields depends on coefficients $A_n$. One usually has that to
large (small) radius of convergence in the real case corresponds
the small (large) one in the $p$-adic case. In order to improve
this situation we shale make an appropriate modification of some
elementary functions. By virtue of the simplicity and enormous
applications in overall theoretical and mathematical physics we
shall concentrate our attention on the exponential, trigonometric,
and hyperbolic functions.

Recall that the series  \bea  \varphi_{\mu,\nu}^\epsilon (x) =
\sum_{n=0}^\infty \epsilon^n \, \frac{x^{\mu n+ \nu}}{(\mu n +\nu)
!} \label{2.2} \eea contains the following functions: $\exp x \,
(\epsilon =1,\, \mu =1, \, \nu =0),\, \, \cos x \,  (\epsilon =
-1,\, \mu =2, \, \nu =0), \, \,  \sin x \, (\epsilon =1,\, \mu =2,
\, \nu =1), \, \, \cosh x \, (\epsilon =1,\, \mu =2, \, \nu =0),$
and $\sinh x \, (\epsilon =1,\, \mu =2, \, \nu =1) .$ It is well
known from classical analysis that series (\ref{2.2}) is
everywhere convergent on ${\mathbb R}$.

{\bf Theorem 1:} Power series \bea  \Phi_{\mu,\nu}^{\epsilon, q}
\, (x ) =  \sum_{n=0}^\infty \epsilon^n \, I_{\mu n +\nu}^{(q)}
\frac{x^{\mu n+ \nu}}{(\mu n +\nu) !} \, , \label{2.3} \eea where
$\epsilon\pm 1, \, 0< q\in {\mathbb Q}, \, \mu \in {\mathbb N}, \,
\nu \in {\mathbb N}_0 = {\mathbb N}\cup \{ 0 \} , $  and \bea
I_{\mu n+ \nu}^{(q)} = \frac{\big( (\mu n+ \nu)! \big)^{\mu n+
\nu}}{q + \big( (\mu n+ \nu)! \big)^{\mu n+ \nu}} \label{2.4} \eea
converges for all $ x\in {\mathbb R}$ and all $x \in {\mathbb
Q}_p$ for every $p$.

{\it Proof:}  In a real case the above theorem follows from the
fact that for large enough $n$ parameter $q$ can be neglected in
comparison to the factorial term and $ I_{\mu n+ \nu}^{(q)}$ may
be approximated by 1. Hence series (\ref{2.3}) asymptotically
behaves like (\ref{2.2}) which is convergent at all real $x$.
Recall \cite{schikhof}  that, in $p$-adic case, a necessary and
sufficient condition for a convergence of (\ref{2.1}) is \bea \mid
A_n x^n\mid_p \rightarrow 0, \quad \mbox{as} \quad n\rightarrow
\infty. \label{2.5} \eea As a consequence  of (\ref{2.5}) it is
enough to consider the $p$-adic norm of the general term in
(\ref{2.3}), i.e.,  \bea   \Big|\epsilon^n  I_{\mu n +\nu}^{(q)}
\frac{x^{\mu n+ \nu}}{(\mu n +\nu) !}\Big|_p  = \frac{|(\mu n
+\nu)!|_p^{\mu n +\nu -1}}{|q+ \big( (\mu n +\nu)! \big)^{\mu n
+\nu}|_p}\, |x|_p^{\mu n +\nu} .\label{2.6}\eea Since the $p$-adic
norm satisfies the strong triangle inequality, one has \bea
 |q+ \big( (\mu n +\nu)! \big)^{\mu n
+\nu}|_p = |q|_p     \label{2.7} \eea for large enough $n$. Note
that \bea  |n!|_p = p^{-(n-n')/(p-1)} , \label{2.8} \eea where
$n'$ is the sum of digits in the canonical expansion of $n$ over
$p$. According to (\ref{2.8}), one has, for the numerator of
(\ref{2.6}),   \bea   \hspace{-2.5cm}  |(\mu n +\nu)! |_p^{\mu n
+\nu -1} \, |x|_p^{\mu n +\nu}  = \big( p^{-\{ [\mu n +\nu - (\mu
n +\nu)']/(p-1) \} [(\mu n +\nu -1)/(\mu n +\nu)]} \, |x|_p
\big)_{n\rightarrow \infty}^{\mu n +\nu} \, \rightarrow 0\,
,\label{2.9} \eea which is valid for any $p$ and all $x\in
{\mathbb Q}_p$ . On the basis of (\ref{2.7})  and (\ref{2.9}) it
follows everywhere convergence on ${\mathbb Q}_p$ for any $p$.
Thus Theorem 1 is proved.

Among all possible examples of analytic functions contained in
power series (\ref{2.3}) we want to point out  the following ones:
$$  \exp_q x  = \sum_{n=0}^{\infty} \frac{(n!)^n}{q + (n!)^n} \, \frac{x^n}{n!}
\eqno(10a)  $$
$$
 \cos_q x  = \sum_{n=0}^{\infty} (-1)^n \, \frac{((2n)!)^{2n}}{q + ((2n)!)^{2n}} \, \frac{x^{2n}}{(2n)!}
\eqno(10b)
$$
$$
 \sin_q x  = \sum_{n=0}^{\infty} (-1)^n \, \frac{((2n +1)!)^{2n +1}}{q + ((2n +1)!)^{2n +1}} \,
 \frac{x^{2n+1}}{(2n+1)!}
\eqno(10c)
$$
$$
 \cosh_q x  = \sum_{n=0}^{\infty}   \frac{((2n)!)^{2n}}{q + ((2n)!)^{2n}} \, \frac{x^{2n}}{(2n)!}
\eqno(10d)
$$
$$
 \sinh_q x  = \sum_{n=0}^{\infty}   \frac{((2n +1)!)^{2n +1}}{q + ((2n +1)!)^{2n +1}} \,
 \frac{x^{2n+1}}{(2n+1)!}
\eqno(10e)
$$

Inverse functions  of (10a)-(10e) can be defined in the usual way,
where coefficients in the power expansions are appropriately
modified. For example,
$$
y = \ln_q x  = \sum_{n=1}^{\infty} (-1)^{n+1} \, a_n^{(q)} \,
\frac{(x - I_0^{(q)})^n}{n} ,
$$
where $I_0^{(q)} = (q+1)^{-1}$ and $a_1^{(q)} = (I_0^{(q)})^{-1} =
q+1, \, \, a_2^{(q)} =I_2^{(q)}/ (I_1^{(q)})^3  = 4 (q+1)^3/(q+4)
, \cdots .$

 \section{ADELIC ASPECTS}

Recall \cite{gelfand} that an adele is an infinite sequence \bea
 a = (a_\infty, \, a_2, \, \cdots , a_p, \, \cdots ) , \label{3.1} \eea
where $a_\infty\in {\mathbb Q}_\infty  =  {\mathbb R} , \, \,
a_p\in {\mathbb Q}_p$ with the restriction that all but a finite
number of $a_p\in {\mathbb Z}_p = \{ x\in {\mathbb Q}_p | \,
|x|_p\leq 1 \} .$ The set of adeles $ {\mathbb A}$ is a ring under
componentwise addition and componentwise multiplication. It is an
additive group $ {\mathbb A}^+$ with respect to addition. The
subset of ${\mathbb A}$  with $\lambda_\infty \neq 0, \, \lambda_p
\neq 0$ for all $p$, and $|\lambda_p|_p = 1$ for all but a finite
number of $p$ is a multiplicative group of ideles ${\mathbb
A}^\ast$. One has a principal adele (idele) if \bea
 r = (r, \, r, \, \cdots , r, \, \cdots ) , \label{3.2} \eea
where $r\in  {\mathbb Q}  \, \, \, (r\in {\mathbb Q}^\ast =
{\mathbb Q}\setminus \{ 0 \})$. One can define a product of norms
on ideles  \bea
 |\lambda| = |\lambda_\infty|_\infty  \, \prod_p |\lambda_p|_p \, , \label{3.3} \eea
where $|\cdot|_\infty$ denotes the usual absolute value. For a
principal idele it yields \bea
 |r| = |r|_\infty  \, \prod_p |r|_p  = 1 . \label{3.4} \eea
Equation (\ref{3.4}) is a well-known product formula for nonzero
rational numbers. An additive character on $ {\mathbb A}^+$ is
\bea \nn \chi_b (a) &=\exp 2\pi i (- a_\infty b_\infty + a_2 b_2 +
\cdots + a_p b_p + \cdots )\\ &
 = \exp (-2\pi i \, a_\infty b_\infty)
\prod_p \exp 2\pi i \, \{ a_p b_p\}_p \, ,\label{3.5}\eea where
$a, b \in  {\mathbb A}^+$, and $\{ x_p\}_p$ denotes a fractional
part of $x_p$. On an idele, \bea \lambda = (\lambda_\infty,
\lambda_2 , \cdots, \lambda_p , \cdots) , \label{3.6}\eea there
exists multiplicative character \bea \hspace{-0.5cm} \pi (\lambda)
= \pi_\infty (\lambda_\infty) \,
 \pi_2 (\lambda_2) \cdots \pi_p (\lambda_p) \cdots\  =
 |\lambda_\infty|_\infty^{c_\infty}  \prod_p |\lambda_p|_p^{c_p} , \label{3.7} \eea
where $c_\infty$ and $c_p$ are complex numbers. Note that in
(\ref{3.5})  and (\ref{3.7}) only finitely many factors are
different from unity.

It may be of physical interest to construct adeles from series
(\ref{2.3}) while their arguments $x$ belong to the principal
adeles (\ref{3.2}).

{\bf Theorem 2:}  Let us have a sequence,  \bea
\Phi_{\mu,\nu}^\epsilon (x)  = \big( \varphi_{\mu,\nu}^\epsilon
(x) ,\, \Phi_{\mu,\nu}^{\epsilon,1/2} (x) , \, \cdots ,
\Phi_{\mu,\nu}^{\epsilon,1/p} (x) , \cdots  \big) , \label{3.8}
\eea where $\varphi_{\mu,\nu}^\epsilon (x)$ is a real series
defined by (\ref{2.2}) ,  and $\Phi_{\mu,\nu}^{\epsilon,1/p} (x)$
is a $p$-adic series defined by (\ref{2.3}) If $x=r \in  {\mathbb
Q}$ then (\ref{3.8}) is an adele.

{\it Proof:} There is no problem with real function
$\varphi_{\mu,\nu}^\epsilon (x)$ for any $x \in {\mathbb Q}$. The
general term of the $p$-adic series (\ref{2.3}) for $q = 1/p $ is
\bea  \epsilon^n \, \frac{\big( (\mu n +\nu)! \big)^{\mu n +\nu
-1}}{1/p + \big( (\mu n +\nu)! \big)^{\mu n +\nu }} \, x^{\mu n +
\nu} = \epsilon^n \, \frac{p\, \big( (\mu n +\nu)! \big)^{\mu n
+\nu -1}}{1 + p\, \big( (\mu n +\nu)! \big)^{\mu n +\nu }} \, x
^{\mu n + \nu}\, . \label{3.9} \eea

It is obvious that $|1+ p\, \big( (\mu n +\nu)! \big)^{\mu n +\nu
}|_p = 1$ . Hence the $p$-adic norm of (\ref{3.9}) is \bea
\frac{1}{p}\big| \big( (\mu n +\nu)! \big)^{\mu n +\nu -1} \big|_p
\, |x|_p^{\mu n +\nu}\, , \label{3.10} \eea which, for a given
$x=r$, can be larger than $1$ only for a finite number of $p$.
Since this conclusion is valid for any $\mu \in {\mathbb N}$ and
$\nu,\, n \in {\mathbb N}_0$ it follows that $
|\Phi_{\mu,\nu}^{\epsilon,1/p} (r)|_p \geq 1 $ only for a finite
number of $p$. So, it is shown  that
$\Phi_{\mu,\nu}^{\epsilon,1/p} (x)$ is an adele when $x$ is a
principal adele.

Note that instead of $\varphi_{\mu,\nu}^{\epsilon} (r)$, which is
$\Phi_{\mu,\nu}^{\epsilon,0} (r)$, one can take for the real term
in (\ref{3.8}) any of the series $\Phi_{\mu,\nu}^{\epsilon,q} (r)$
defined by (\ref{2.3}). It is easy to see that $1/p$ in
(\ref{3.8}) can be substituted by $p^{-s}$, where $s\in {\mathbb
N}$.

\section{ON SUMMATION}

It is not clear that there does exist $0\neq x {\mathbb Q}$ for
which series (\ref{2.3}) is a rational number. By an analogy with
the real case of series (\ref{2.2}) one can expect that there is
not such a possibility. For a trivial case, i.e. $x=0$, one has
\bea  \Phi_{\mu,\nu}^{\epsilon, q} (0) =  \left\{
\begin{array}{ll}
                 0,   &   \nu  \geq 1,  \\
                 \frac{1}{q+1},   &   \nu = 0 .
                 \end{array}    \right.
 \label{4.1} \eea

We shall show that starting from series (\ref{2.3}) one can obtain
a sum of the corresponding functional series.

{\bf Theorem 3:}  The summation formula, \bea &\hspace{-2cm} \nn
\sum_{n=0}^\infty
 \big( (\mu n +\nu)! \big)^{\mu n +\nu -1} \, x^{\mu n} \Big\{
 \frac{[\big( \mu (n+ 1) +\nu \big)!]^\mu \,  (\mu n +\nu +1)_\mu^{\mu n +\nu
-1}}{q + [\big( \mu (n +1)+\nu \big)!]^{\mu (n+1)+\nu}} \, x^\mu  \\
& - \frac{1}{q+ \big( (\mu n +\nu)! \big)^{\mu n +\nu }}\Big\} =
-\frac{(\nu!)^{\nu - 1}}{q+ (\nu!)^\nu} \, ,\label{4.2} \eea where
$(\mu n +\nu + 1)_\mu = (\mu n +\nu + 1)\, (\mu n +\nu + 2) \cdots
(\mu n +\nu + \mu)$, has a place for all $0\neq x \in {\mathbb R}$
as well as for all $0\neq x \in {\mathbb Q}_p$ for every $p$.

{\it Proof:} Expansion  (\ref{2.3}) of $\Phi_{\mu,\nu}^{-1, q}
(x)$ can be rewritten as follows: \bea & \hspace{-2cm} \nn
\Phi_{\mu,\nu}^{-1, q} (x) = \frac{(\nu !)^{\nu-1}}{q + (\nu
!)^\nu} \, x^\nu +  \big( (\mu +\nu)! \big)^{\mu  +\nu -1} \,
x^{\mu +\nu}  \Big[ \frac{\big( (2 \mu  +\nu)! \big)^\mu \,  (\mu
+\nu +1)_\mu^{\mu +\nu -1} }{q + \big( (2 \mu  +\nu)! \big)^{2 \mu
+\nu}} \, x^\mu \\ & - \frac{1}{q + \big( (\mu  +\nu)! \big)^{\mu
+\nu }} \Big] + \cdots \, , \label{4.3} \eea \bea &
\hspace{-2.3cm} \nn -\Phi_{\mu,\nu}^{-1, q} (x) = (\nu !)^{\nu -
1} \, x^\nu \, \Big[\frac{\big( (\mu  +\nu)! \big)^\mu \, (\nu
+1)_\mu^{\nu-1}}{q+ \big( (\mu  +\nu)! \big)^{\mu  +\nu }}\, x^\mu
- \frac{1}{q + (\nu!)^\nu} \Big] + \big( (2 \mu  +\nu)! \big)^{2
\mu  +\nu -1} \\ & \hspace{-1.5cm} \times  x^{2 \mu +\nu} \,
\Big[\frac{\big( ( 3\mu +\nu)! \big)^\mu \, (2 \mu +\nu +1)_\mu^{2
\mu +\nu-1}}{q+ \big( (3 \mu +\nu)! \big)^{3 \mu  +\nu }}\, x^\mu
- \frac{1}{q + \big( (2 \mu +\nu)! \big)^{2 \mu  +\nu }} \Big] +
\cdots . \label{4.4} \eea Addition of (\ref{4.3}) and (\ref{4.4})
with division by $x^\nu$ for $x\neq 0$ leads to formula
(\ref{4.2}).

In the particular case $(\mu = 1,\, \nu = 0, \, x=-1, \, q=1)$, we
have  \bea    \sum_{n=0}^\infty (-1)^n  \, \frac{\big( (n + 1)!
\big)^n \, \big[ 1+ (2+n) (n!)^n  \big] + (n!)^{n-1}}{\big[1+
(n!)^n \big]\, \big[1+ \big((n+1)!\big)^{n+1} \big]} = \frac{1}{2}
,\label{4.5} \eea which is the result valid in ${\mathbb R}$  and
all ${\mathbb Q}_P$.

\section{DISCUSSION AND CONCLUDING REMARKS}

The first question that we want to discuss is related to the
possible applications of the power series (\ref{2.3}) . Recall
that the interest in $p$-adic models is mainly motivated by some
indications \cite{volovich1} that space-time at the Planck scale
should be analyzed using $p$-adic mathematics. According to this
point of view, let us consider classical cosmological solutions of
the Einstein gravitational equations. These equations for the
scale factor $R (t)$ of the homogeneous and isotropic universe are
\bea  \frac{\ddot{R} (t)}{R (t)} = -\frac{\kappa\, (\rho + 3
p)}{6} \, , \quad \left(  \frac{\dot{R}}{R} \right)^2 +
\frac{k}{R^2} = \frac{\kappa\, \rho}{3}\, , \label{5.1} \eea where
$\rho$ is the energy density and $p$ is the corresponding pressure
$(\kappa = 8\pi G; \, k= +1, \, -1,\, 0)$. Assume that $\rho$  and
$p$ depend on time in the way that enables an application of
(\ref{2.3}) for $R(t)$. Among the three cases let us choose the
following one:  \bea \nn k = 0\, ,\\   \nn \rho_q (t) =
\frac{3}{\kappa} \left[ \frac{d}{dt} \ln (\exp_q H t) \right]^2\,
,\\   p_q (t) = - \rho_q (t) - \frac{2}{\kappa}\, \frac{d^2}{dt^2}
\ln (\exp_q H t) \, , \label{5.2} \\ \nn R_q (t)  = H^{-1} \exp_q
Ht , \quad H = \sqrt {\Lambda/3} . \eea An analogous situation is
for $k =1$ with $R (t) = H^{-1} \cosh_q Ht$ and $k = - 1$ with $R
(t) = H^{-1} \sinh_q H t$. If $\kappa \in {\mathbb Q}$ all theses
models may be treated either real  or $p$-adic. By decreasing
parameter $q$, expansion of the universe given by (\ref{5.2}) can
be done arbitrary close to the de Sitter model. Namely, when
$q\rightarrow 0: \, \rho_q \rightarrow \Lambda/\kappa, \, \,
p_q\rightarrow - \Lambda/\kappa ,$  and $R_q (t) \rightarrow R (t)
= \sqrt{3/\Lambda} \exp \sqrt{\Lambda/3}\, t$, where $\Lambda$ is
the cosmological constant. It would be interesting to find a
scalar-field model that leads to $\rho_q (t)$ and $p_q (t)$. A
classical cosmological solution for $k=+1$ and $R_q (t) =
\sqrt{3/\Lambda} \cosh_q \sqrt{\Lambda/3}\, t$ can be further used
in $p$-adic quantum cosmology \cite{dragovich2}, which is a
generalization of the Hartle-Hawking approach to the wave function
of the universe.

Let us suppose that the parameter $q$ is a quotient $(q =
l_{Pl}/l)$ of the Planck length ($l_{Pl}\sim 10^{-33}$ cm) and a
length that characterizes the given scale. For example, the
unification length in the electroweak theory is about $10^{-17}$cm
and for the GUT one gets $l\sim 10^{-29}$cm.  In such a way only
at the Planck scale parameter $q \, \, (q=1)$ cannot be neglected.
So, $q =1$ in the real case and $q = 1/p$ in the $p$-adic case
should be natural values in the high-energy limit ($E \sim
10^{19}$ GeV) . Taking $\Phi_{\mu,\nu}^{\epsilon, 1} (x)$ instead
of $\varphi_{\mu,\nu}^{\epsilon} (x)$ in (\ref{3.8}), one has a
unification at the Planck scale of real and $p$-adic functions
(\ref{2.3}) in the form of adeles.

Parameter $q$ regularizes (\ref{2.2}) to enlarge the region of
convergence from $|x|_p < 1, \, p\neq 2\, $  (and $|x|_2 <
\frac{1}{2}$) to the whole ${\mathbb Q}_p \, $.  In the limit
$q\rightarrow 0$ the regularized functions tend to the usual ones.


\bigskip
\bigskip

\bigskip

\begin{thebibliography}{99}


\bibitem{volovich1} I.V. Volovich, Class. Quantum Grav. {\bf 4}
L83 (1987); B. Grossman, Phys. Lett. B {\bf 197}, 101  (1987):
P.G.O. Freund and M. Olson, Phys. Lett. B {\bf 199}, 186 (1987);
P.G.O. Freund and E. Witten, {\it ibid} {\bf 199}, 191 (1987);
P.H. Frampton, Y. Okada, and M.R. Ubriaco, {\it ibid} {\bf 213},
260 (1988); I.Ya Aref'eva, B.G. Dragovich and I.V. Volovich, {\it
ibid} {\bf 209}, 445 (1988): {\bf 212}, 283 (1988); {\bf 214} 339
(1988); L.O. Chekhov and A.Yu. Zinoviev, Commun. Math. Phys. {\bf
130}, 623 (1990); P.G.O. Freund, J. Math. Phys. {\bf 33}, 1148
(1992).

\bibitem{volovich2} C. Alacoque, P. Ruelle, E. Thiran, D.
Verstegen, and J. Weyers, Phys. Lett. B {\bf 211}, 59 (1988); V.S.
Vladimirov and I.V. Volovich, Commun. Math. Phys. {\bf 123}, 659
(1989); B.L. Spokoiny, Phys. Lett. B {\bf 221}, 120 (1989); Y.
Meurice, Int. J. Mod. Phys. A {\bf 4}, 5133 (1989); E.I. Zelenov,
J. Math. Phys. {\bf 32}, 147 (1991); A.Yu. Khrennikov, {\it ibid}
{\bf 32}, 932 (1991).

\bibitem{roth} B.D.B. Roth, Phys. Lett. B {\bf 213}, 263 (1988);
E. Melzer, Int. J. Mod. Phys. A {\bf 4}, 4877 (1989); V.A.
Smirnov, Mod. Phys. Lett. A {\bf 6}, 1421 (1991); M.D. Missarov,
Phys. Lett. B {\bf 272}, 36 (1991).

\bibitem{dragovich1} B.G. Dragovich, P.H. Frampton, and B.V.
Urosevic, Mod. Phys. Lett. A {\bf 5}, 1521 (1990); B.G. Dragovich,
{\it ibid} {\bf 6}, 2301 (1991).

\bibitem{dragovich2} I.Ya. Aref'eva, B.G. Dragovich, P.H.
Frampton, and I.V. Volovich, Int. J. Mod. Phys. A {\bf 6}, 4341
(1991).

\bibitem{zelenov} E.I. Zelenov, J.Math. Phys. {\bf 33}, 178
(1992); A.Yu. Khrennikov, J. Math. Phys. {\bf 33}, 1636 (1992).

\bibitem{dragovich3} I.Ya. Aref'eva, B.G. Dragovich, and I.V.
Volovich,  Phys. Lett. B {\bf 200}, 512 (1988); B.G. Dragovich,
Phys. Lett. B {\bf 256}, 392 (1991); On factorial perturbation
series, preprint No. If-91-011.

\bibitem{schikhof} W.H. Schikhof, {\it Ultrametric Calculus} (Cambridge U.P., Cambridge,
1984).

\bibitem{mahler} K. Mahler, {\it $p$-Adic Numbers and Their Functions
}(Cambridge U.P., Cambridge, 1981).

\bibitem{gelfand}  I.M. Gel'fand, M.I. Graev, and I.I.
Piatetskii-Shapiro, {\it Representation Theory and Automorphic
Functions} (Nauka, Moscow, 1966).





\end{thebibliography}
\end{document}